\documentclass[letterpaper,twocolumn,fleqn]{article} 

\usepackage{ist}
\usepackage{algorithm}
\usepackage{array}
\usepackage{amsmath}
\usepackage{amssymb}
\usepackage[utf8]{inputenc}
\usepackage[export]{adjustbox}
\usepackage{textcomp}
\usepackage{threeparttable}
\usepackage{stfloats}
\usepackage{csquotes}
\usepackage[super]{nth}
\usepackage{caption}
\usepackage{url}
\usepackage{verbatim}
\usepackage{siunitx}
\usepackage[font={footnotesize}, textfont=sf]{subfig} 
\usepackage[table,xcdraw]{xcolor}
\usepackage{fourier}
\usepackage{color}
\usepackage{multirow}
\usepackage{graphicx}
\usepackage{url}
\usepackage{amsmath}
\usepackage{wrapfig}
\usepackage{booktabs}
\usepackage{threeparttable}
\usepackage[citebordercolor=white, linkbordercolor=white, urlbordercolor=black, hidelinks, pdfborderstyle={/S/U/W 1}, breaklinks]{hyperref}
\usepackage{cleveref}

\usepackage[T1]{fontenc}
\usepackage{times}

\title{SOLAS: Superpositioning an Optical Lens in Automotive Simulation}

\author{Daniel~Jakab~$^{1,}$$^5$, Julian~Barthel~$^2$, Alexander~Braun~$^2$, Reenu~Mohandas~$^{1,}$$^5$, Brian~Michael~Deegan~$^{3,}$$^5$, Mahendar~Kumbham~$^4$, Dara~Molloy~$^{3,4}$, {Fiachra}~Collins~$^4$, Anthony~Scanlan~$^1$, Ciarán~Eising~$^{1,}$$^5$\\
$^1$Dept. of Electronic and Computer Engineering, University of Limerick, Castletroy, Co. Limerick  V94 T9PX, Ireland\\
$^2$Faculty of Electrical Engineering \& Information Technology, University of Applied Sciences, Düsseldorf 40476, Germany\\
$^3$Dept. of Electrical and Electronic Engineering, University of Galway, Galway H91 TK33, Ireland\\
$^4$Valeo Vision Systems, Tuam, Galway DY1 22DJ, Ireland\\
$^5$Lero the Science Foundation Ireland Research Centre for Software, University of Limerick, Limerick, V94 T9PX, Ireland\\}

\date{} 

\begin{document} 

\maketitle 

\thispagestyle{empty} 


\begin{abstract}
Automotive Simulation is a potentially cost-effective strategy to identify and test corner case scenarios in automotive perception. Recent work has shown a significant shift in creating realistic synthetic data for road traffic scenarios using a video graphics engine. However, a gap exists in modeling realistic optical aberrations associated with cameras in automotive simulation. This paper builds on the concept from existing literature to model optical degradations in simulated environments using the Python-based ray-tracing library {\em KrakenOS}. As a novel pipeline, we degrade automotive fisheye simulation using an optical doublet with +/-\ang{2} Field of View(FOV), introducing realistic optical artifacts into two simulation images from SynWoodscape and Parallel Domain Woodscape. We evaluate {\em KrakenOS} by calculating the Root Mean Square Error (RMSE), which averaged around 0.023 across the RGB light spectrum compared to Ansys Zemax OpticStudio, an industrial benchmark for optical design and simulation. Lastly, we measure the image sharpness of the degraded simulation using the ISO12233:2023 Slanted Edge Method and show how both qualitative and measured results indicate the extent of the spatial variation in image sharpness from the periphery to the center of the degradations.
\end{abstract}

\section{Introduction}
\label{sec:intro}
Automotive Simulation is seen as a potential strategy for creating corner case training data in computer vision. One of the key challenges in automotive simulation is the increasing need to standardize simulations and create reliable data for training purposes. A recent survey~\cite{jakab2024surround} has shown that automotive simulators lack the optical lens from the real world, which is crucial for computer vision. Without modeling the realistic artifacts of the optical lens in simulation, the generated synthetic data is incomparable in quality to that of the real world. In recent work, there is a clear trend in simulating optical lenses on both synthetic and realistic data in the form of a Point Spread Function (PSF) map or grid where the optical artifacts are convolved with images \cite{yang2023image,yang2023aberration, Wittpahl2018, lehmann2018resolution}. One of the difficulties of modeling an optical lens on realistic data is the challenge of deconvolving the original image with the PSFs of the original lens used to capture the image \cite{lehmann2018resolution, heide2013high}. This is undoubtedly a bottleneck when it comes to lens simulation. In many cases, the recorded realistic data in public datasets does not have this information and can only be estimated or measured in most cases \cite{yang2023aberration,lehmann2018resolution, heide2013high, jakab2024measuring}. As a means of addressing this gap in research, we utilize the Pythonic ray-tracing library, KrakenOS~\cite{herrera2022krakenos} to degrade synthetic fisheye images intrinsically warped from a video graphics engine. We have chosen an optical doublet for our experiments to investigate the open-source tool, KrakenOS~\cite{herrera2022krakenos}, which has never been used extensively for optical simulations to the authors' knowledge. The optical doublet, a simplistic optical model, is used as a system under test to investigate the capabilities of KrakenOS~\cite{herrera2022krakenos}.
The main contributions are as follows:
\begin{itemize}
    \item \textbf{A performance comparison between KrakenOS~\cite{herrera2022krakenos} and Ansys Zemax OpticStudio software~\cite{ansysZemaxOpticStudio2024}}, an industrial benchmark for optical design and simulation.
    \item \textbf{A novel optical degradation pipeline} for SynWoodscape~\cite{sekkat2022synwoodscape} and Parallel Domain Woodscape~\cite{paralleldomains2023}.
    \item \textbf{Investigation of degraded automotive fisheye simulations} for the presence of optical artifacts.
    \item \textbf{Measurement for image sharpness} in the baseline and degraded fisheye automotive simulations using the ISO12233:2023~\cite{ISOBS2023} standard.
\end{itemize}

\section{Related Work}
Recent work has shown that training a computer vision neural network without accounting for the camera lens is insufficient for perception systems, especially when autonomous vehicles have diverse cameras deployed, such as wide-angle cameras for low-speed maneuvering and automated parking \cite{Eising2021, Yogamani2019}. As a means of a cost-effective strategy, there have been attempts to replicate the real-world environment as seen through a camera lens where geometric distortion is introduced using the intrinsic properties of a real-world lens into simulation \cite{ sekkat2022synwoodscape, paralleldomains2023}. However, geometric distortion is only one optical artifact of many where it is important to model other optical artifacts in simulation.
It has been shown in a previous study by Heide et al. \cite{heide2013high} that optical artifacts are spatially variant which is shown from the patchwise-estimated PSFs of two simple thin lenses, one being that of a biconvex lens at f/2.0 and the second a plano-convex lens at f/4.5. Computer vision tasks, especially depth estimation are impacted by the spatial variance of optical artifacts \cite{jakab2024surround, yang2023aberration, chang2019deep}. By accounting for these artifacts, especially chromatic aberration, it has been shown that monocular depth estimation models improve performance on real-world tasks \cite{chang2019deep}. This spatially varying aspect of optical lenses is especially challenging to simulate in a wide FOV perspective. It must be accounted for in fisheye optical systems where strong radial distortion affects these artifacts. As shown in Yang et al. \cite{yang2023aberration}, optical aberrations can be simulated by ray-tracing the optical system. We can fine-tune computer vision systems to cameras of interest by introducing optical lenses into our training. 
\begin{figure*}[t]
    \centering
    \vspace{-20pt}
    \captionsetup[subfloat]{captionskip=-70pt}
    \subfloat{\includegraphics[width=6in, keepaspectratio]{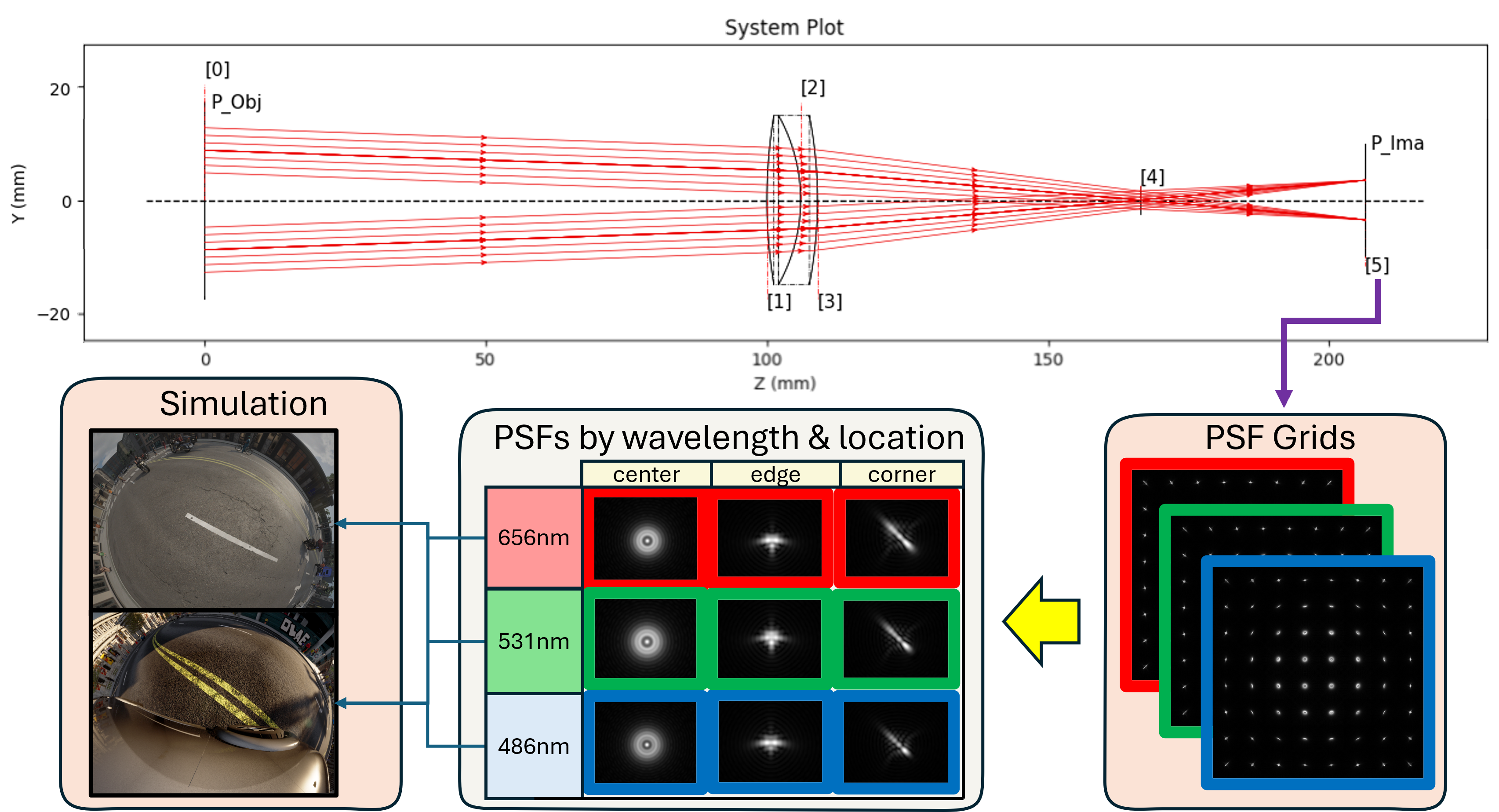}}
    \caption{\textbf{\textit{Degradation Pipeline of an Optical Doublet where all surfaces are labeled between [0-5] with ray-tracing performed between +/-\ang{2} FOV and light wavelengths of $w=(656nm, 531nm, 486nm)$ (red, green, blue). The generated PSF grid outputs are then convolved with two simulation images.}}}
    \label{fig:doub}
    \vspace{-0pt}
\end{figure*}
\begin{table}[]
\centering
\caption{Properties and surfaces of the optical doublet lens.}
\label{tab:doublet-spec}
\resizebox{\columnwidth}{!}{%
\begin{tabular}{|l|l|l|l|l|l|}
\hline
\rowcolor[HTML]{C0C0C0} 
No. & Name:         & Material: & Thickness:& Diameter: & Radius Curvature: \\ \hline
0   & $P_{Obj}$        & AIR       & 100.000   & 30.000    & 0.000             \\
1   & Surface 1     & BK7       & 6.000     & 30.000    & 92.847            \\
2   & Surface 2     & F2        & 3.000     & 30.000    & -30.716           \\
3   & Surface 3     & AIR       & 57.376    & 30.000    & -78.197           \\
4   & Aperture Stop & AIR       & 39.496    & 5.000     & 0.000             \\
5   & $P_{Ima}$        & AIR       & 0.000     & 20.000    & 0.000             \\ \hline
\end{tabular}%
}
\vspace*{-\baselineskip}
\end{table}
\begin{figure*}[t]
    \centering
    \subfloat[\label{fig:psf-grid-deg-8x6}]{\includegraphics[width=3.1in, keepaspectratio]{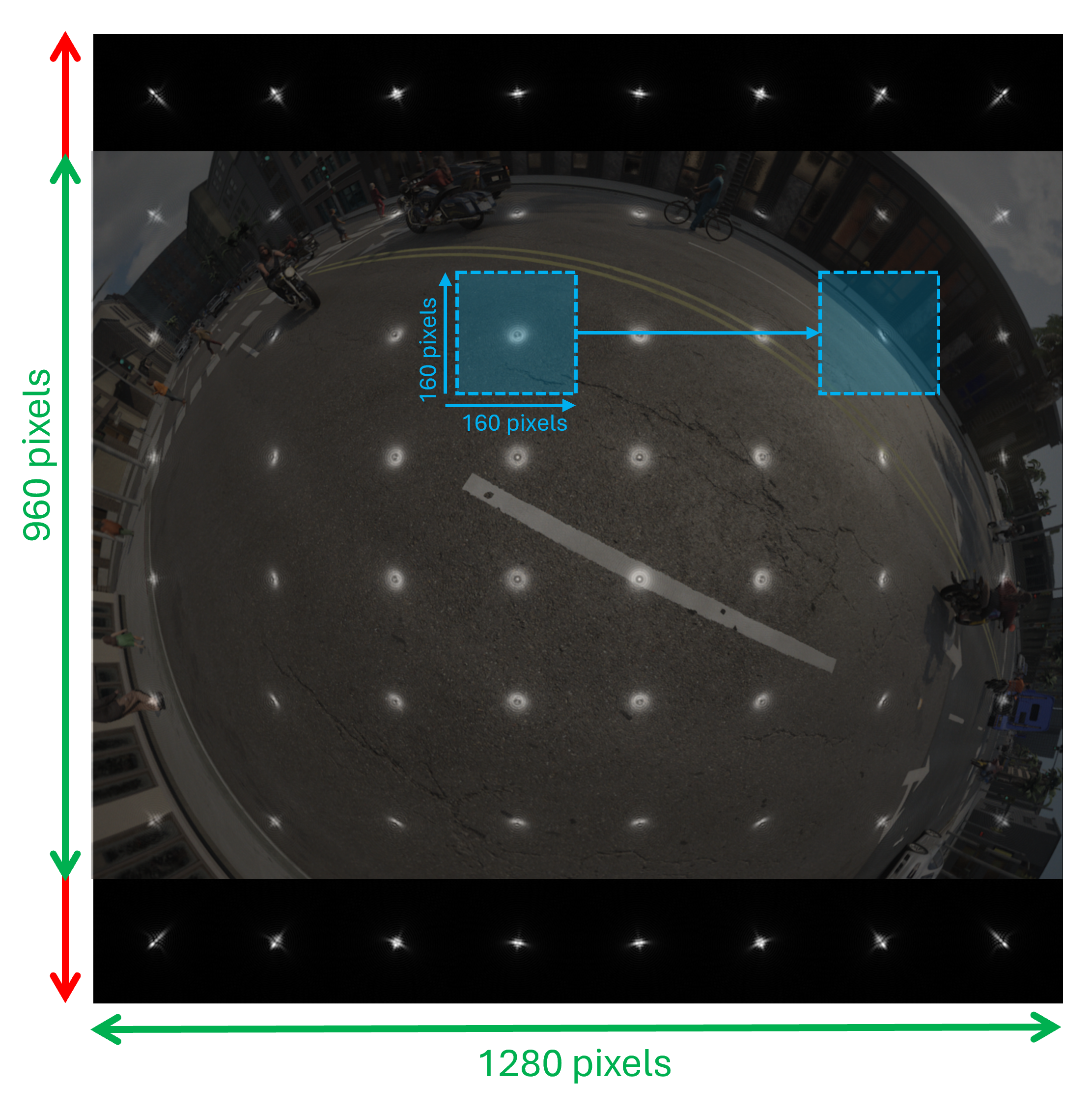}}
    \qquad
    \subfloat[\label{fig:psf-grid-deg-32x24}]{\includegraphics[width=3.1in, keepaspectratio]{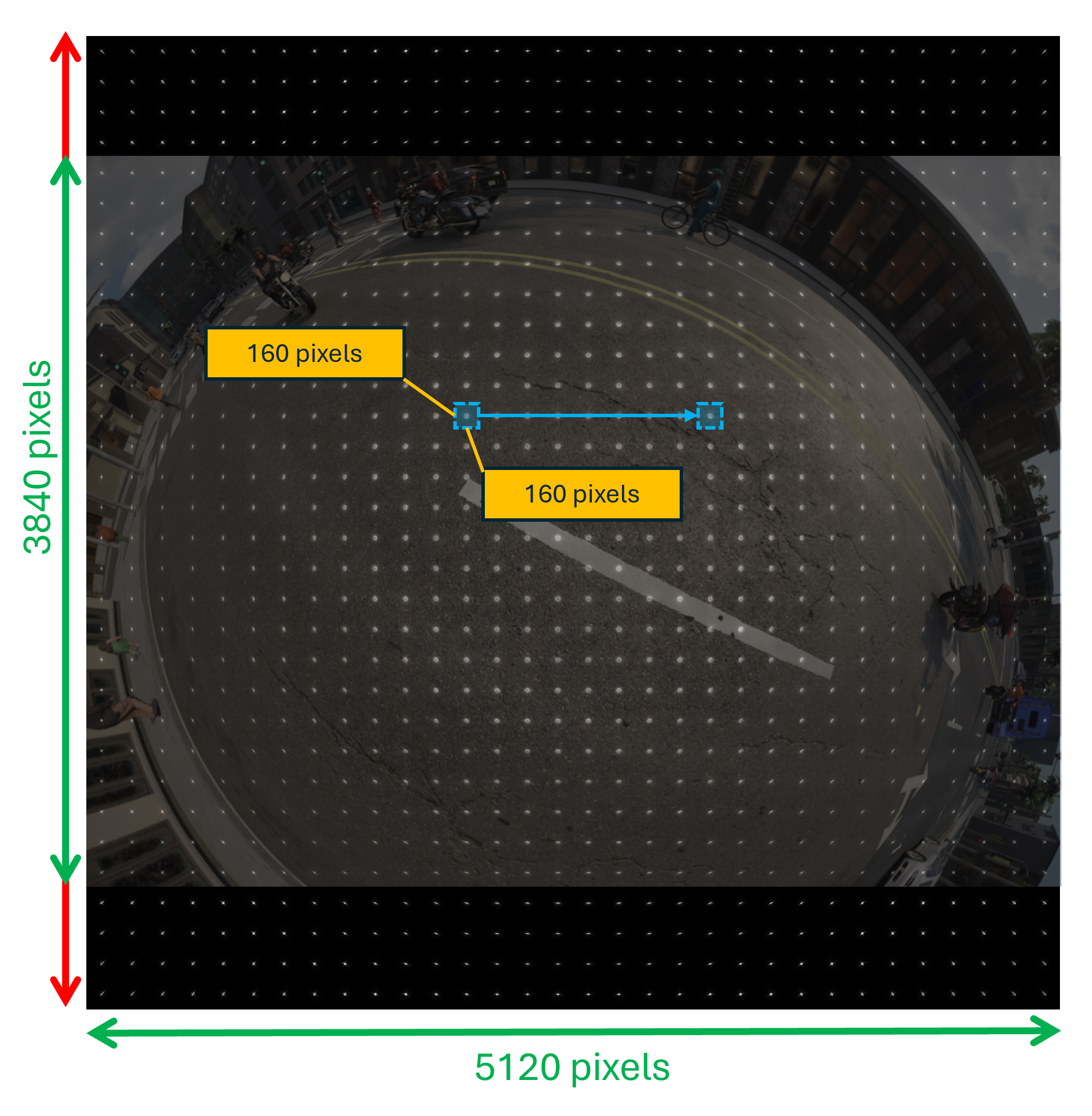}}
    \caption{\emph{\textbf{Convolution of PSF grids using both (a) $8\times8$ and (b) $32\times32$ grid sizes with $160\times160$ pixels per PSF each.}}}.
    \label{fig:psf-grid-deg}
\end{figure*}
\begin{figure*}[t]
    \centering
    \subfloat{\includegraphics[width=5.8in, keepaspectratio]{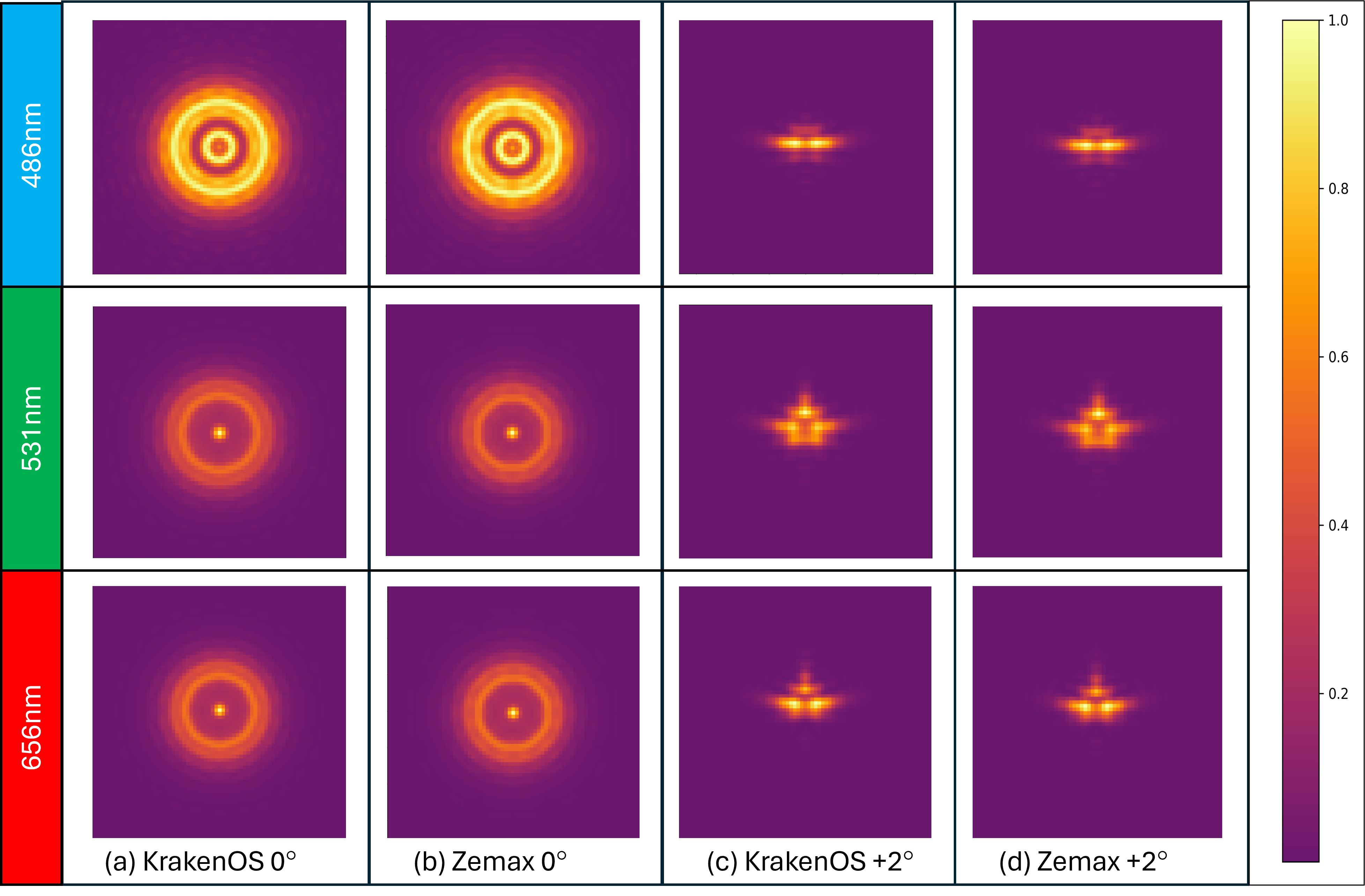}}
    \caption{\emph{\textbf{Fraunhofer Point Spread Functions at \ang{0} and $+\ang{2}$ FOV where $w\in[486nm, 531nm, 656nm]$}}}.
    \label{fig:psf-comp}
\end{figure*}
\section{Methodology}
As a strategy for implementing realistic optical degradations in imagery, we propose the following approach:
\begin{enumerate}
    \item Build an optical doublet in KrakenOS~\cite{herrera2022krakenos}. Figure \ref{fig:doub} shows a two-dimensional view of the System Under Test (SUT) containing six surfaces specified as:
    \begin{itemize}
        \item{\textbf{Object Plane ($P_{Obj}$):} A plane from which parallel rays appear at a given angle to the optical axis.}
        \item \textbf{Surface 1:} BK7 glass material with refractive index $n\approx 1.517$ (according to a light wavelength of 588nm).
        \item \textbf{Surface 2:} F2 glass material with refractive index $n \approx 1.62$ (according to a light wavelength of 588nm).
        \item \textbf{Surface 3:} The back of the lens with a negative radius of curvature.
        \item \textbf{Aperture Stop:} Limits the amount of light that passes through the system.
        \item \textbf{Image Plane ($P_{Ima}$):} Plane at which light rays converge to form an image.
    \end{itemize}
    \item Build and compare the same optical system in Ansys Zemax OpticStudio~\cite{ansysZemaxOpticStudio2024}, a well-established industrial tool for optical design and simulations. PSF analysis is used to compare and fine-tune the KrakenOS~\cite{herrera2022krakenos} design to produce a performance equivalent to that of Ansys Zemax OpticStudio~\cite{ansysZemaxOpticStudio2024}.
    \item Simulate the optical system between a range of +/-\ang{2} FOV. Two different simulation approaches were used:
    \begin{enumerate}
        \item Generate $8\times8$ PSF grids for red, green, and blue (RGB) where the respective wavelengths are $w_{r} = 656nm$, $w_{g} = 531nm$ and $w_{b} = 486nm$, respectively. Figure \ref{fig:doub} shows the top-level procedure of extracting the PSF grids from the system, each with their respective light wavelengths. The sizes of the PSF grids are $1280\times1280$ pixels each, where each generated PSF is of size $160\times160$ pixels.
        \item Generate $32\times32$ PSF grids for RGB simulated at a large scale for a resolution of $5120\times5120$ where the same PSF size is used, i.e., $160\times160$ pixels. Note: the same wavelengths are used as before. This provides a finer degree of degradation.
    \end{enumerate}
    \item Convolve the PSF grids on two $1280\times960$ fisheye simulations on the Mirror View Left (MVL position). An image was chosen of one camera position from both the SynWoodscape \cite{sekkat2022synwoodscape}, and Parallel Domain Woodscape \cite{paralleldomains2023} datasets to introduce spatially variant optical artifacts in the fisheye images. The objective is to degrade baseline images according to the spatial position of each pixel to the PSF grids. Due to the rectangular nature of the simulation images, the PSF grids were resized to $8\times6$ and $32\times24$ PSF grids, respectively. As shown in Figure \ref{fig:psf-grid-deg}, two different grid sizes were used to test the quality of the degradation and observe the qualitative change in PSF shape both at a scale where the PSFs are of the same size as the resolution and at a size where the regions of convolution are four times smaller than the size of the PSFs across the spatial domain. Investigations have found that having a finer convolution with a greater number of PSFs, as in the case of the $32\times24$ PSF grids, produced a smoother degradation across the image from the center to the periphery. The two degradation strategies that are used are as follows:
    \begin{enumerate}
        \item The $8\times6$ PSF RGB grids are convolved with its spatially equivalent region in the simulation images where each PSF has a square region of $160\times160$ pixels (see Figure \ref{fig:psf-grid-deg-8x6}).
        \item The $32\times24$ PSF RGB grids are convolved where each 
        PSF has a square region of $160\times160$ pixels (i.e., the same size as before), providing a much finer degradation strategy. Due to the large number of PSFs, the PSF grid size increases and is four times larger. Each PSF is convolved by a $40\times40$ region of the simulation image. This is indicated by the 1280:5120 or 1:4 grid ratio in Figure \ref{fig:psf-grid-deg-32x24}.
    \end{enumerate}
    \item Investigate image sharpness across the spatial region of the degraded fisheye images using the ISO12233:2023 Slanted Edge Method from previous work \cite{jakab2024measuring, van2023tool}.
\end{enumerate}
\section{Build \& Finetune KrakenOS}
This work's major objective is to establish a pipeline and optically degrade simulation images where degraded simulation would be seen through a real lens. In Herrera et al ~\cite{herrera2022krakenos}, KrakenOS, an open-source python-based ray-tracing tool, provides a relatively new research alternative for optical simulation having a performance difference of $9.0\times10^{-8}$mm compared to Ansys Zemax OpticStudio~\cite{ansysZemaxOpticStudio2024}, a standard commercial tool for optical simulation design~\cite{herrera2022krakenos}. In general, we wish to understand how the open-source tool KrakenOS~\cite{herrera2022krakenos} is different in terms of optical simulation compared to commercially used tools. This paper compares PSF outputs between KrakenOS~\cite{herrera2022krakenos} and Ansys Zemax OpticStudio~\cite{ansysZemaxOpticStudio2024} for further analysis. A direct comparison in generated PSFs is crucial for further insight into the performance difference between both tools. To ascertain that the difference is relatively small, KrakenOS~\cite{herrera2022krakenos} is utilized and fine-tuned to reflect the optical outputs of Ansys Zemax OpticStudio~\cite{ansysZemaxOpticStudio2024} using the parameters in Table \ref{tab:doublet-spec}. Figure \ref{fig:psf-comp} illustrates the comparison of KrakenOS~\cite{herrera2022krakenos} and Ansys Zemax OpticStudio ~\cite{ansysZemaxOpticStudio2024} Fraunhofer PSFs between \ang{0} and +\ang{2} FOV, varying the wavelength of light between red, green, and blue (RGB). The RGB wavelengths are chosen to reflect the RGB nature of the colored simulation images used for degradation~\cite{sekkat2022synwoodscape, paralleldomains2023}. In Figure \ref{fig:psf-comp}, there are similarities where, in both tools, the red and green wavelengths of light produce PSFs that are Gaussian-like in nature where the center of the PSF has the highest value shown as a bright dot at \ang{0} FOV (see Figure \ref{fig:psf-comp} (a) and (b) for wavelengths 656nm and 531nm).
\begin{table}[t]
\centering
\footnotesize
\caption{Root Mean Square Error(RMSE) for KrakenOS generated PSFs compared to Ansys Zemax OpticStudio~\cite{ansysZemaxOpticStudio2024} where $\theta_{y}$ represents the change in the angle of FOV along the y-axis.}
\label{tab:rmse-results}
\begin{tabular}{lll}
\hline
\rowcolor[HTML]{C0C0C0} 
\textbf{$\theta_{y}$(FOV)} & \textbf{Wavelength(nm)} & \textbf{RMSE}  \\ \hline
0               & 486                     & 0.042          \\
0               & 531                     & \textbf{0.020} \\
0               & 656                     & 0.021          \\ \hline
-2              & 486                     & 0.022          \\
-2              & 531                     & \textbf{0.020} \\
-2              & 656                     & 0.021          \\ \hline
+2               & 486                     & 0.022          \\
+2               & 531                     & \textbf{0.020} \\
+2               & 656                     & 0.022          \\ \hline
\end{tabular}%
\begin{tablenotes}
  \small
  \item Note: the green wavelength at 531nm produced the lowest RMSE out of all three RGB wavelengths.
\end{tablenotes}
\vspace*{-15pt}
\end{table}
This would represent an in-focus system automatically optimized in Ansys Zemax OpticStudio~\cite{ansysZemaxOpticStudio2024}, representing the highest performance possible from this optical system. However, for the blue wavelength (i.e., 486nm), the PSFs at \ang{0} FOV show slightly stronger rings than for the other two wavelengths, indicating that for this particular design setup, images with blue characteristics would have the lowest optical performance out of all cases, with some degree of astigmatism and spherical aberration.
As a further step, the same optical model was constructed in Ansys Zemax OpticStudio~\cite{ansysZemaxOpticStudio2024} where the PSFs were simulated between +/-\ang{2} FOV along the y-axis (see Figure \ref{fig:psf-comp} (c) and (d) for the +\ang{2} FOV cases). To quantitatively compare both results, the Root Mean Square Error(RMSE) was calculated where KrakenOS~\cite{herrera2022krakenos} is compared to Ansys Zemax OpticStudio~\cite{ansysZemaxOpticStudio2024} results. In Table \ref{tab:rmse-results}, an RMSE of around 0.02 was obtained in most cases for all three wavelengths. The blue wavelength has a performance difference of twice that of the other wavelengths. This may be due to the non-gaussian or spherical aberration in the blue wavelength at the center of the FOV, which can pose a greater challenge for open-source comparison. The small difference in results between both tools demonstrates that KrakenOS~\cite{herrera2022krakenos} can match industrial performance.
\begin{figure*}[t]
    \centering
    \subfloat[\label{fig:KrakenOS-SW-deg}]{\includegraphics[width=2.8in, keepaspectratio]{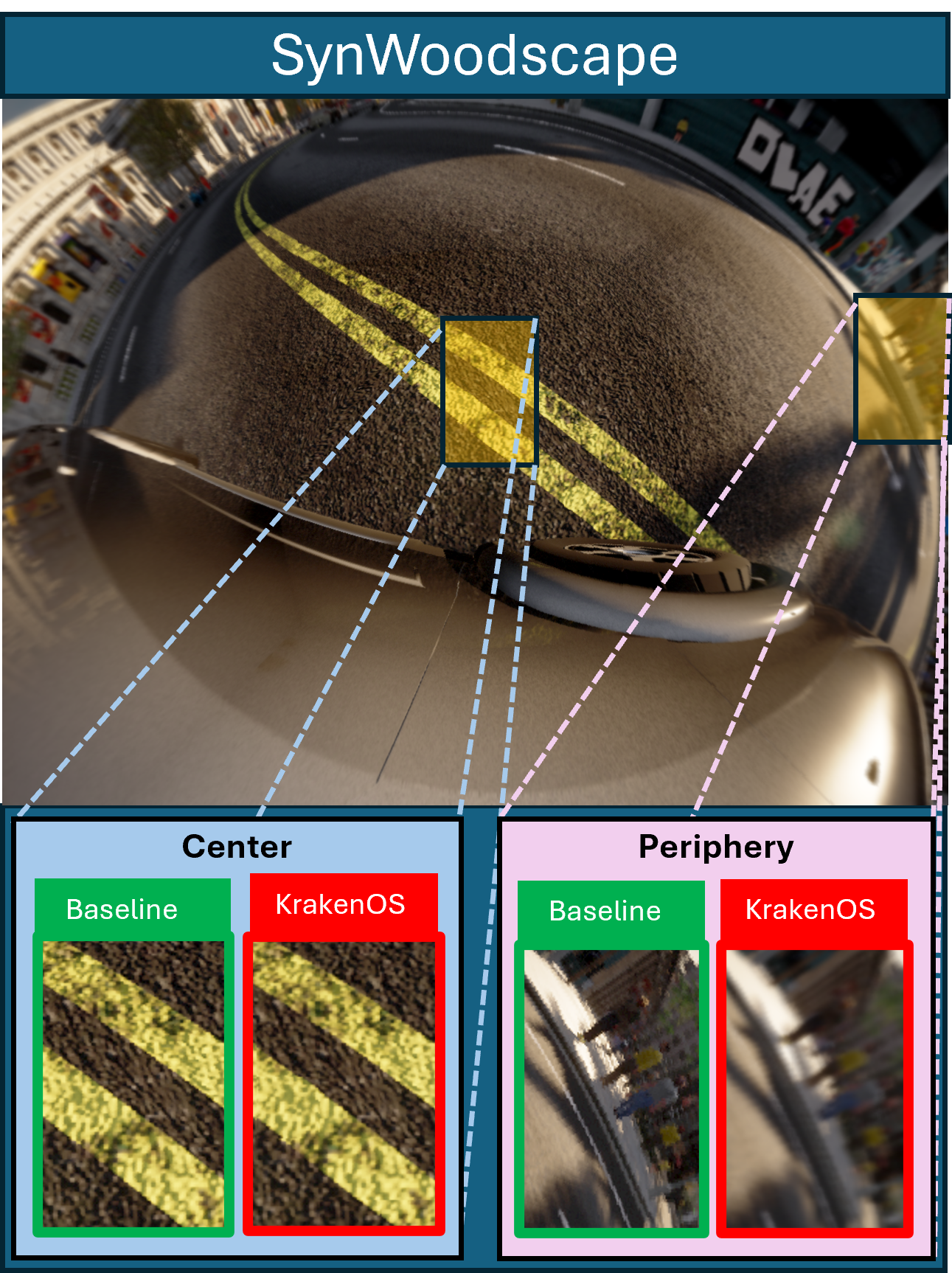}}
    \qquad
    \subfloat[\label{fig:KrakenOS-PDW-deg}]{\includegraphics[width=2.82in, keepaspectratio]{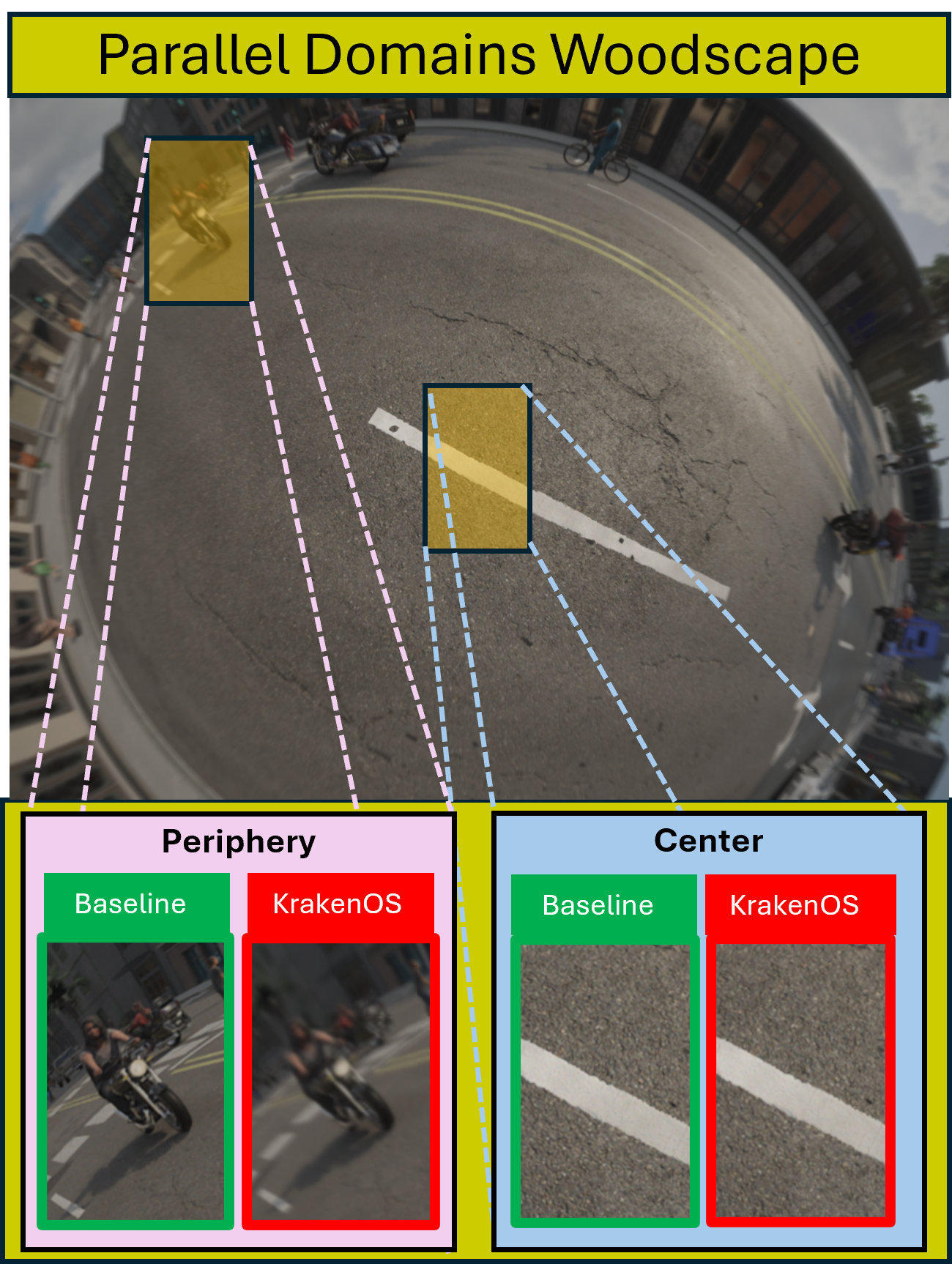}}
    \caption[\label{fig:KrakenOS-deg}]{\emph{\textbf{Optical degradations of both (a) SynWoodscape and (b) Parallel Domain Woodscape where green frames represent the baseline cases with no degradations applied and red frames represent degraded cases after convolving the test images with KrakenOS.}}}.
    \label{fig:sim-degrade}
\end{figure*}

\section{Optical Degradations of Fisheye Automotive Simulation}
In the previous section, we demonstrated a side-by-side comparison of Ansys Zemax OpticStudio~\cite{ansysZemaxOpticStudio2024} with KrakenOS~\cite{herrera2022krakenos}, our chosen method for optical simulations in this work. Using the KrakenOS~\cite{herrera2022krakenos} toolkit, we have established a strategy of degrading the simulated images by extracting the PSF outputs of the optical doublet for the RGB wavelengths of light and convolving them with the simulation images. In this section, we will present and discuss the results of this convolution qualitatively and by measuring the average edge contrast, otherwise known as the sharpness of the simulations before and after degradation is applied. 

In Figure \ref{fig:sim-degrade}, both the SynWoodscape and Parallel Domain Woodscape Mirror View Left(MVL) images are shown where two regions of interest are shown towards the periphery of the images. 
When qualitatively comparing both simulations, SynWoodscape has a lower-quality image with aliasing along the road and also on the pedestrians. In contrast, Parallel Domain Woodscape has much smoother edges and qualitatively shows more realistic lighting conditions than SynWoodscape. 

Once the optical doublet is applied, in both simulation images, a distinct blur can be seen in the same regions of interest, indicating the presence of optical artifacts in the simulations, such as coma and astigmatism towards the periphery of the images (see Figure \ref{fig:sim-degrade}). This is expected from the PSF grids where the comet-shaped PSFs are skewed according to the lens design giving rise to optical artifacts. Previous work has shown that applying a Gaussian blur filter uniformly degrades images by reducing contrast and removing information from the image~\cite{jakab2024ss}. This work shows that the blur evident from optical degradations is not uniform but spatially variant (see Figure \ref{fig:KrakenOS-SW-deg} and Figure \ref{fig:KrakenOS-PDW-deg} for the center and periphery regions). In addition, optical degradations would carry many non-gaussian degradations, such as chromatic aberration. In the optical doublet used in this work, the degraded images from Figure \ref{fig:sim-degrade} do not qualitatively show any evidence of this effect, and more investigation is needed. Chromatic aberration is typically difficult to correct in camera systems especially fisheye due to extreme radial distortions and field curvatures \cite{jakab2024surround}.

Qualitatively, results show that the optical doublet, a key lens component of optical systems (including fisheye), can be replaced for more complex optical systems in our pipeline. A caveat to our experiments would be that the simulation images used in this work already have the geometric properties of the fisheye lens initially used from the Woodscape dataset \cite{sekkat2022synwoodscape, Yogamani2019}. The degradation itself is not realistic because the warped simulation is convolved with a completely different lens. Simulations should be recorded, warped, and convolved with the same lens to capture realistic degradations with our pipeline. 
\begin{figure*}[t]
    \centering
    \subfloat[\label{fig:ardan-ss-sfr}]{\includegraphics[width=3.3in, keepaspectratio]{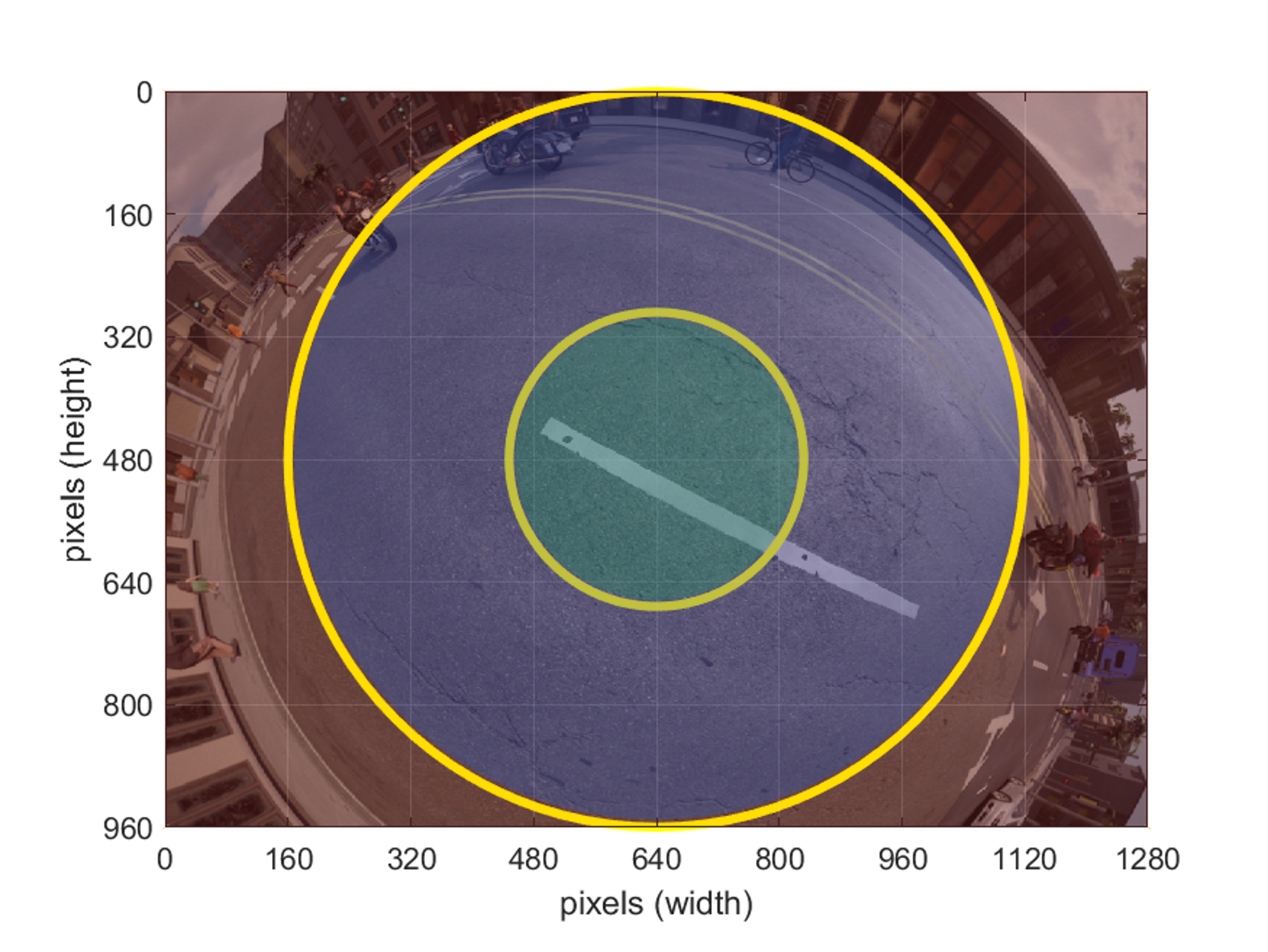}}
    \hspace{0.2cm}
    \subfloat[\label{fig:ardan-ss-sfr-mtf50}]{\includegraphics[width=3.3in, keepaspectratio]{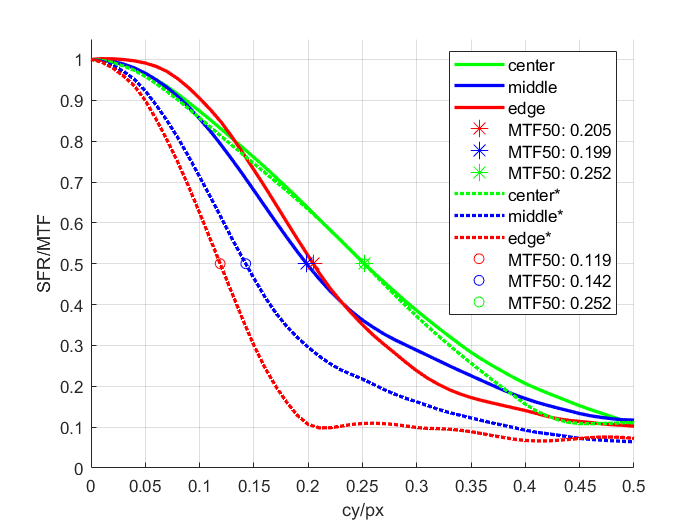}}
    
    \caption{\emph{\textbf{Mean MTF50 measurement using the SS-SFR technique where (a) depicts the SS-SFR categorization using radial annuli and (b) the MTF measurements of automotive simulation images before and after degradation. Note: The solid lines are baseline, and the dashed lines degrade measurements once the optical doublet is applied.}}}
    \label{fig:ss-sfr-comp}
\end{figure*}

\section{Measuring the Contrast of Degraded Fisheye Automotive Simulations}
To further evaluate the degraded automotive simulations, we measure 50\% of the Modulation Transfer Function(MTF50) curve, a typical measure for camera quality and image sharpness (otherwise known as the image contrast). We utilize the Synthetic Scenes Spatial Frequency Response(SS-SFR) from previous work \cite{jakab2024ss} to effectively measure and isolate valid horizontal slanted edges from both fisheye simulation scenes. The centroid locations of the slanted edges are categorized into three circular radii determined in previous work~\cite{jakab2024measuring}(see Figure \ref{fig:ardan-ss-sfr} where the radii are mapped onto one simulation image).
As shown in Figure \ref{fig:ardan-ss-sfr-mtf50}, the MTF50 values are measured where the curves drop to 50\% of their value. A general observation shows that the range of results varies between 0.2-0.25cy/px before degradation is applied. The images' central regions return the highest image contrast out of all three regions, where other regions are tightly clustered around 0.2cy/px. Both middle and edge regions return roughly the same value of results. In contrast, once degradation is applied, both middle and central regions drop in image sharpness, where the middle region varies by -0.057cy/px and the edge region degraded by -0.086cy/px. This is a significant drop and also shows that the periphery of the simulations is affected more by coma and astigmatism than any other regions, which is expected. Interestingly, the central region returns the same result before and after degradation. This shows that the optical system preserves image contrast in the central regions of the automotive simulations.
\section{Limitations \& Future Work}
A few observations can be made about this pipeline:
\vspace{-5pt}
\begin{itemize}
    \item Further analysis comparing open-source KrakenOS~\cite{herrera2022krakenos} and the industrial tool Ansys Zemax OpticStudio~\cite{ansysZemaxOpticStudio2024} is needed.
    \item Future work must adapt to wide-angle optical systems such that the realism of fisheye simulations can be improved.
    \item KrakenOS~\cite{herrera2022krakenos} uses the Central Processing Unit (CPU), where computational time can be improved by offloading calculations to the Graphics Processing Unit (GPU) instead.
\end{itemize}

\section{Conclusion}
In this work, we have successfully extracted the properties of an optical doublet lens using KrakenOS, a Python-based ray-tracing library. These properties were extracted as three PSF grids reflecting the RGB color space. We convolved the generated PSF grids for RGB with two fisheye automotive simulations. We also calculated the RMSE between KrakenOS and Ansys Zemax OpticStudio where RMSE averaged around 0.02 across the entire FOV. Image contrast degraded for both the edge and middle regions of the simulations but remained the same for the center, where the simulated lens is spatially variant. 



\bibliographystyle{IEEEtran}
\small
\bibliography{ieee-references}
\section{Acknowledgments} 
This work was supported by the Science Foundation Ireland grant 13/RC/2094\_2. The authors would like to thank Ben McKeon, a Mathematics PhD student at the University of Limerick, for suggesting the Python-based KrakenOS optical tool for this study. His advice made these experiments possible.

\end{document}